\documentclass[journal]{IEEEtran}
\usepackage[final]{graphicx}
\DeclareGraphicsExtensions{.eps}

\usepackage[cmex10]{amsmath}

\usepackage{amsmath,epsfig,bm,amssymb,amsthm}
\usepackage{psfrag,accents}
\usepackage{cite}
\usepackage{color}
\ifCLASSINFOpdf
\else
\fi

\usepackage{amsmath,epsfig,bm,amssymb,amsthm,balance}

\newcommand{\sinc}{{\mathrm {sinc}}}
\newcommand{\et}{{\mathrm {e}}}
\newcommand{\PEP}{{\mathrm {PEP}}}
\newcommand{\rt}{{\mathrm {r}}}
\newcommand{\SR}{{\mathrm {SR}}}
\newcommand{\mf}{{\mathrm {mf}}}
\newcommand{\RiD}{{\mathrm {R}_i\mathrm{D}}}

\newcommand{\cp}{{\mathrm {cp}}}
\newcommand{\IDFT}{{\mathrm {IDFT}}}
\newcommand{\DFT}{{\mathrm {DFT}}}

\newcommand{\Zc}{\mathcal{Z}}

\newcommand{\Hb}{\mathbf{H}}

\newcommand{\Sb}{\mathbf{S}}

\newcommand{\Vb}{\mathbf{V}}

\newcommand{\s}{\mathbf{s}}

\newcommand{\0}{\mathbf{0}}

\newcommand{\x}{\mathbf{x}}

\newcommand{\y}{\mathbf{y}}

\newcommand{\w}{\mathbf{w}}
\newcommand{\CN}{\mathcal{CN}}
\newcommand{\Vc}{\mathcal{V}}

\newcommand{\I}{\mathbf{I}}

\newcommand{\C}{\mathbf{C}}
\newcommand{\G}{\mathbf{G}}
\newcommand{\B}{\mathbf{B}}

\newcommand{\tG}{\widetilde{G}}
\newcommand{\tl}{\tilde{l}}

\newcommand{\Cbc}{\bm{\mathcal{C}}}
\newcommand{\psib}{\bm{\psi}}
\newcommand{\phib}{\bm{\phi}}
\newcommand{\modN}[1]{\langle {{#1}} \rangle_N}
\newcommand{\intv}{{\mathrm {intv}}}

\IEEEoverridecommandlockouts

\begin{document}


\title{Differential Distributed Space-Time Coding with Imperfect Synchronization in Frequency-Selective Channels}

\author{
	\IEEEauthorblockN{M. R. Avendi and Hamid Jafarkhani, {\it Fellow, IEEE}  \\
Email: \{m.avendi, hamidj\}@uci.edu
\thanks{The authors are with the Center for Pervasive Communications and Computing, University of California, Irvine, USA. This work was supported in part by the NSF Award CCF-1218771.
}
}
}

\maketitle

\begin{abstract}
\label{abs}
Differential distributed space-time coding (D-DSTC) is a cooperative transmission technique that can improve diversity in wireless relay networks in the absence of channel information. Conventionally, it is assumed that channels are flat-fading and relays are perfectly synchronized at the symbol level. However, due to the delay spread in broadband systems and the distributed nature of relay networks, these assumptions may be violated. Hence, inter-symbol interference (ISI) may appear. This paper proposes a new differential encoding and decoding process for D-DSTC systems with multiple relays over slow \emph{frequency-selective} fading channels with imperfect synchronization. The proposed method overcomes the ISI caused by frequency-selectivity and is robust against synchronization errors while not requiring any channel information at the relays and destination. Moreover, the maximum possible diversity with a decoding complexity similar to that of the conventional D-DSTC is attained. Simulation results are provided to show the performance of the proposed method in various scenarios.
\end{abstract}


\IEEEpeerreviewmaketitle

\section{Introduction}
\label{sec:intro}
Cooperative communication techniques make use of the fact that, since users in a network can listen to a source during its transmission phase, they will be able to re-broadcast the received data to the destination in another phase. Therefore, the overall diversity and performance of a network would benefit from a virtual antenna array that is constructed cooperatively by multiple users.

Distributed space-time coding (DSTC) is a cooperative transmission technique for wireless relay networks. In DSTC networks \cite{DSTC-HJ,DSTC-Laneman,DSTC-Y,DSTC-Kaveh}, the relays cooperate to combine and forward the received symbols to construct a space-time block code \cite{stc-vt-hj} at the destination. Coherent detection of transmitted symbols can be achieved by providing the instantaneous channel state information (CSI) of all transmission links at the destination. Although this requirement can be accomplished by sending pilot (training) signals and using channel estimation techniques, it is not feasible or efficient in relay channels as there are more channels involved. Moreover, the computational complexity and overhead of channel estimation increase proportionally with the number of relays. Also, all channel estimation techniques are subject to impairments that would directly translate to performance degradation. 

The above issues motivate the use of non-coherent detection in relay networks. When no CSI is available at the relays and destination, differential DSTC (D-DSTC) over flat-fading channels has been studied in \cite{D-DSTC-Y,D-DSTC-Amin,D-DSTC-Giannakis}. D-DSTC only needs the second-order statistics of the channels at the relays and provides many of the advantages of multiple antenna deferential modulation systems in \cite{diff-vt-hj,dqostc-hj}. For example, the constructed space-time block code at the destination together with differential encoding provides the opportunity to apply non-coherent detection without any CSI.

In broadband systems, signals experience \emph{frequency-selective} fading. Frequency-selective channels are modeled by finite-impulse response (FIR) filters with independent coefficients/taps in the baseband. As such, the received signal is the convolution between the transmitted signal and the FIR channel. Clearly, this causes inter-symbol-interference (ISI). Orthogonal-frequency division multiplexing (OFDM) is widely used as a remedy for the ISI in frequency-selective channels in both single-antenna and MIMO systems \cite{gold_wireless,MIMO-OFDM-Gia,MIMO-OFDM-Paulraj}. However, channel estimation is still required for coherent detection \cite{CE-OFDM1,CE-OFDM2}. To avoid channel estimation in frequency-selective channels, differential OFDM (D-OFDM) was developed for single-antenna and MIMO systems in \cite{OFDM-Luise,OFDM-Kaiser,MIMO-OFDM-Liu,MIMO-OFDM-Borgman,MIMO-OFDM-Pauli,MIMO-OFDM-Himsoon,sftc-hj}. In cooperative communications over frequency-selective channels, the number of channels grows proportionally with the number of relays and the number of taps in the FIR filter. Thus, the importance of developing differential OFDM schemes becomes even more significant in relay networks over broadband systems.

On the other hand, because of the distributed nature of relay networks, signals from different relays may be received at the destination asynchronously with different delays. Such a delay difference is not necessarily an integer multiple of the symbol duration. This so-called synchronization error between relays causes ISI as well. To overcome this ISI both time-domain and frequency-domain approaches have been considered in the literature. For coherent DSTC with one/two relays over flat-fading channels, time-domain approaches to synchronization errors have been studied in  \cite{ADSTC-Valenti,ADSTC-Hua,ADSTC-HJ,ADSTC-Olafsson}. These methods require both the CSI of all transmission links and the amount of relative synchronization error between relays. On the other hand \cite{ADSTC-FF-2R-Xia1,ADSTC-FF-2R-Xia2} consider a frequency-domain approach using OFDM for networks with two relays over flat-fading channels using coherent detection. The method has been extended to networks with multiple relays over flat-fading channels in \cite{ADSTC-Rajan}. Moreover, \cite{ADSTC-FS-2R-Xia,ADSTC-FS-2R-Yoon} develop the OFDM approach for networks with two relays over frequency-selective channels using coherent detection. However, \cite{ADSTC-FF-2R-Xia1,ADSTC-FF-2R-Xia2,ADSTC-FS-2R-Xia,ADSTC-FS-2R-Yoon,ADSTC-Rajan} assume that the synchronization errors are integer multiples of the symbol duration.

Typically the received signal at the destination is passed through a matched-filter and sampled at the symbol rate. It is desired to capture the peak value of the received signal, while the contribution of the filter side-lobes are zero or very small. However, due to the fractional synchronization errors, the peak values of signals will be drifted from the sampling point and the side-lobes contribute to the sampled signal. Consequently, this causes another source of ISI while the received signal-to-noise ratio (SNR) is degraded proportional to the synchronization errors and the matched-filter function. In \cite{async-lulu1,asynch-lulu2,asynch-lulu3} a time-domain approach using over-sampling has been developed for two-way relay networks using coherent detection to combat fractional synchronization errors.
In \cite{ADSTC-HJ}, a time-domain over-sampling method is utilized for the coherent DSTC. These methods assume a perfect knowledge of the synchronization delays and channels at the receiver. Also, in \cite{ADSTC-FF-us} we studied a frequency-domain approach for D-DSTC systems with two relays over flat-fading channels.

In this paper, a differential encoding and decoding process together with an OFDM approach is designed. The method combats the ISI caused by both frequency-selective channels and synchronization errors in relay networks. We assume arbitrary delays that are not necessarily integer multiples of the symbol duration. We consider the case that a source communicates with a destination via multiple relays. Neither CSI nor synchronization errors are available at the destination. We propose a new sampling method, called double sampling, to combat the effects of synchronization errors. The new double sampling method does not assume any knowledge of synchronization errors at the destination. It is shown that the scheme significantly improves system performance when fractional synchronization errors exist. For uncoded systems, the achievable diversity approaches the number of relays at high transmit power. Moreover, channel coding and interleaving are integrated to exploit the intrinsic frequency diversity of frequency-selective channels. Simulation results show the effectiveness of the proposed method in various scenarios.

The outline of the paper is as follows: Section \ref{sec:D-OFDM} describes the channel model, encoding and decoding processes, the new double sampling scheme and channel coding and interleaving. Simulation results are provided in Section \ref{sec:Sim}. Section \ref{sec:con} concludes the paper.

\emph{Notations:} Bold lower and upper-case letters denote vector and matrix, respectively. $(\cdot)^t$, $(\cdot)^*$ and $(\cdot)^H$ denote transpose, complex conjugate and transpose conjugate of a vector or matrix, respectively. $|\cdot|$ and $\| \cdot \|$ stand for absolute value of a complex number and the Euclidean norm of a vector, respectively. $\I_R$ and $\0$ are the $R \times R$ identity and zero matrices, respectively. $\CN(\0,\sigma^2 \I_R)$ stands for a circularly symmetric Gaussian random vector with zero mean and covariance $\sigma^2 \I_R$. $\mbox{E}\{\cdot\}$ denotes the expectation. Both $\et^{(\cdot)}$ and $\exp(\cdot)$ stand for the exponential function. $\Zc$ is the set of integer numbers. $\sinc(x)=\sin(\pi x)/(\pi x)$. $j=\sqrt{-1}.$

\begin{figure}[t]
\psfrag {Source} [] [] [1.0] {Source}
\psfrag {Relay1} [] [] [1.0] {Relay 1}
\psfrag {Relay2} [] [] [1.0] {Relay 2}
\psfrag {RelayR} [] [] [1.0] {Relay R}
\psfrag {Destination} [] [] [1.0] {Destination}
\psfrag {s1} [r] [] [1.0] {}
\psfrag {x1} [r] [] [1.0] {}
\psfrag {x2} [r] [] [1.0] {}
\psfrag {y1} [r] [] [1.0] {}
\psfrag {f1} [r] [] [1.0] {$\{q_{1,l}\}$}
\psfrag {f2} [bl] [] [1.0] {$\{q_{2,l}\}$}
\psfrag {fR} [tl] [] [1.0] {$\{q_{R,l}\}$}
\psfrag {g1} [l] [] [1.0] {$\{g_{1,l}\}$}
\psfrag {g2} [l] [] [1.0] {$\{g_{2,l}\}$}
\psfrag {gR} [l] [] [1.0] {$\{g_{R,l}\}$}
\centerline{\epsfig{figure={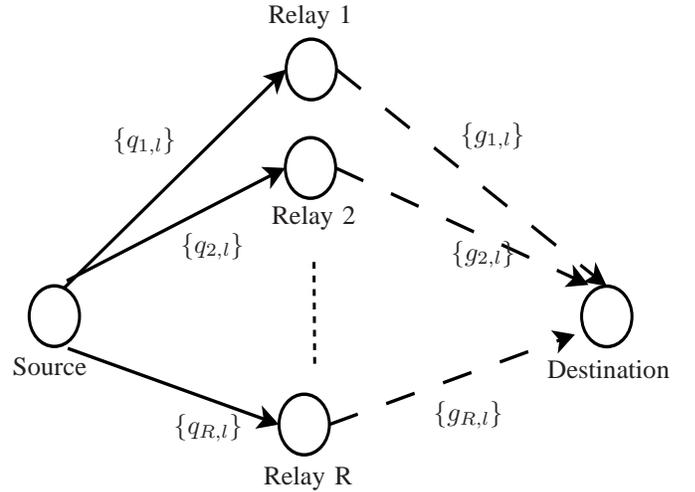},height=6.5cm,width=8.5cm}}
\caption{Cooperative network under consideration, the source communicates with the destination through $R$ relays.}
\label{fig:sysmodel}
\end{figure}

\begin{figure*}[t]
\psfrag {PtS} [] [] [1][-90] {Parallel to Serial}
\psfrag {v1} [] [] [1] {$\{v_1[n]\}$}
\psfrag {v2} [] [] [1] {$\{v_R[n]\}$}
\psfrag {V} [] [] [1] {$\{\Vb[n]\}$}
\psfrag {Unitary} [] [] [1] {Unitary}
\psfrag {STC} [] [] [1] {Matrix}
\psfrag {Diff} [] [] [1] {Differential}
\psfrag {Encod} [] [] [1] {Encoding}
\psfrag {IDFT} [] [] [1] {IDFT}
\psfrag {Add} [] [] [1] {Add}
\psfrag {Pulse} [] [] [1] {Pulse}
\psfrag {Shape} [] [] [1] {Shape}
\psfrag {CP} [] [] [1] {$N_{\cp_1}$}
\psfrag {s1} [] [] [1] {$\{s_1[n]\}$}
\psfrag {s2} [] [] [1] {$\{s_R[n]\}$}
\psfrag {S1} [] [] [1] {$\;\; \{S_1[m]\}$}
\psfrag {S2} [] [] [1] {$\{S_R[m]\}$}
\psfrag {S1cp} [] [] [1] {$$}
\psfrag {S2cp} [] [] [1] {$$}
\centerline{\epsfig{figure={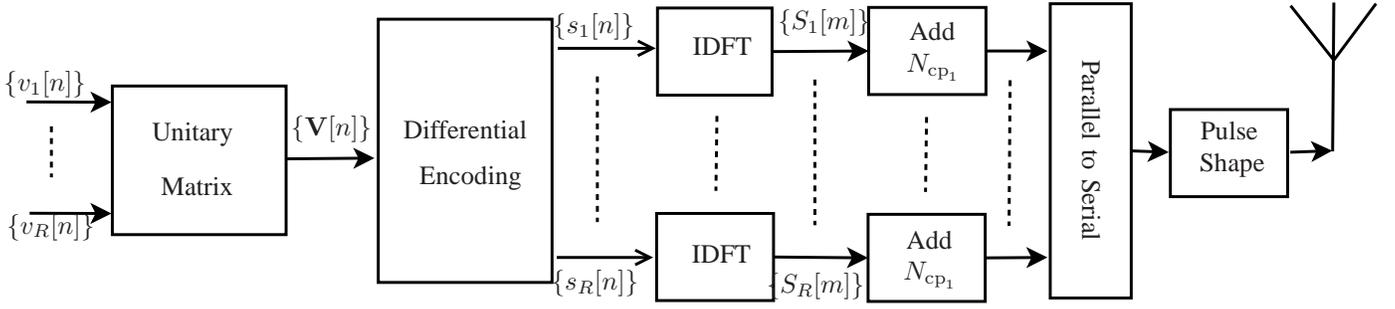},width=18cm}}
\caption{Encoding process at the source, for $n,m=0,\cdots,N-1$.}
\label{fig:sourceblk}
\end{figure*}

\section{Differential OFDM DSTC}
\label{sec:D-OFDM}
The wireless relay network under consideration is depicted in Fig.~\ref{fig:sysmodel}. In the network, there is one source, $R$ relays and one destination. All nodes have one antenna and transmission is half-duplex (i.e., each node can only transmit or receive at any time). A two-phase transmission process is employed. In Phase I, the source transmits a stream of symbols to the relays. In Phase II, the relays configure the received signals, amplify and forward them to the destination. We assume slow frequency-selective fading channels. A frequency-selective fading channel has $L$ independent delay paths with arbitrary power delay profiles. The discrete-time baseband equivalent channels is modeled by an $L-$tap FIR filter with transfer function 
\begin{equation}
\label{eq:h[n]}
h[n]=\sum \limits_{l=0}^{L-1} h_l \delta[n-l],
\end{equation}
where $n$ is the discrete-time index and $h_l$ is the $l$-th path channel coefficient.
If a circular sequence $\{x[n]\},\; n=-L,\cdots,0,\cdots, N-1$ is the input of the channel, the channel output is expressed by \cite{gold_wireless}
\begin{equation}
y[n]=\sum \limits_{l=0}^{L-1} h_l x[\modN{n-l}]=h[n] \otimes x[n],
\end{equation}
\label{eq:y_fs}
for $n=0,\cdots,N-1$, where $\modN{\cdot}$ represents the modulo $N$ operation and $\otimes$ is the circular convolution operation.

The discrete-time baseband equivalent channel from the source to the $i$th relay ($\SR_i$) and from the $i$th to the destination ($\RiD$) are represented by $L-$tap FIR filters with transfer functions 
\begin{equation}
\label{eq:qg[n]}
\begin{split}
q_i[n]&=\sum \limits_{l=0}^{L-1} q_{i,l} \delta[n-l],\\
g_i[n]&=\sum \limits_{l=0}^{L-1} g_{i,l} \delta[n-l]
\end{split}
\end{equation}
where $q_{i,l}\sim \CN(0,\sigma_{q_{i,l}}^2)$ and $g_{i,l}\sim \CN(0,\sigma_{g_{i,l}}^2)$ are the multipath channel coefficients of $\SR_i$ and $\RiD$, respectively, for $i=1,\cdots,R$. Also, the power of $L$ paths are normalized such that $\sum \limits_{l=0}^{L-1} \sigma_{q_{i,l}}^2=1$ and $\sum \limits_{l=0}^{L-1} \sigma_{g_{i,l}}^2=1$.


In the next subsections, the proposed DSTC system, capable of dealing with frequency-selective channels and synchronization errors, is described in detail. The method combines differential encoding and decoding with an OFDM approach and is referred to as differential OFDM (D-OFDM) DSTC. 

\subsection{Encoding Process at Source}
As mentioned, the transmission process is divided into two phases. In Phase I, the source encodes information bits and transmits them to the relays. The encoding process at the source is depicted in Fig.~\ref{fig:sourceblk} and is described as follows. 

First, the information bits are converted to modulation (or data) symbols from constellation set $\Vc$ (such as PSK, QAM). Depending on the number of relays and the type of constellation, appropriate $R \times R$ unitary matrices $\Cbc=\{\Vb|\Vb^H\Vb=\Vb\Vb^H=\I_R \}$ are designed. 
The set of $R$ data symbols  $\{ v_1[n],\cdots,v_R[n] \}\in \Vc$ are then encoded to unitary matrices $\Vb[n]\in \Cbc$ for $ n=0,\cdots, N-1$. For example, for a network with $R=2$ relays, the $2 \times 2$ orthogonal design (OD) can be employed as \cite{DSTC-HJ,D-DSTC-Y}
\begin{equation}
\label{eq:alamouti}
\Vb[n]= \frac{1}{\sqrt{|v_1[n]|^2+|v_2[n]|^2}}
\begin{bmatrix}
v_1[n] & -v_2^*[n] \\
v_2[n] & v_1^*[n]
\end{bmatrix}, 
\end{equation}
to encode two data symbols $v_1[n],v_2[n]$ into one data matrix $\Vb[n]$ or equivalently $2N$ data symbols into $N$ data matrices for $n=0,\cdots, N-1$.

Next, data matrices $\{\Vb[n]\}$ are differentially encoded as \cite{diff-vt-hj}
\begin{equation}
\label{eq:sk-ofdm}
\begin{split}
\s[n]^{(k)}=\Vb[n]^{(k)} \s[n]^{(k-1)}
\end{split}
\end{equation}
for $n=0,\cdots,N-1$ to obtain $N$ vectors of length $R$. Let us denote 
$\s[n]^{(k)}=\left[ s_{1}[n]^{(k)}, \cdots, s_{R}[n]^{(k)} \right]^t$. Here, $(k)$ is the block-index which will be omitted whenever there is no confusion. Also, for $k=1$, the reference vector is defined as $\s[n]^{(0)}=[\;1 \; 0 \cdots 0\;]^t$ with length $R$. Note that $\s^H[n] \s[n]=1$. 

\begin{figure*}[t]
\psfrag {X1} [b] [] [1] {$\{X_{i,1}[m]\}$}
\psfrag {X2} [t] [] [1] {$\{X_{i,R}[m]\}$}
\psfrag {X1cp} [] [] [1] {}
\psfrag {X2cp} [] [] [1] {}
\psfrag {R1} [] [] [1] {$\{Z_{i,1}[m]\}$}
\psfrag {R2} [] [] [1] {$\{Z_{i,R}[m]\}$}
\psfrag {CP2} [] [] [1] {$N_{\cp_2}$}
\psfrag {CP1} [] [] [1] {$N_{\cp_1}$}
\psfrag {Add} [] [] [1] {Add}
\psfrag {Remove} [] [] [1] {Remove}
\psfrag {Pulse} [] [] [1] {Pulse}
\psfrag {Shape} [] [] [1] {Shape}
\psfrag {PtS} [] [] [1][-90] {Parallel to Serial}
\psfrag {StP} [] [] [1][-90] {Serial to Parallel}
\psfrag {stc} [] [] [1] {DSTC}
\psfrag {configure} [] [] [1] {Configure }
\centerline{\epsfig{figure={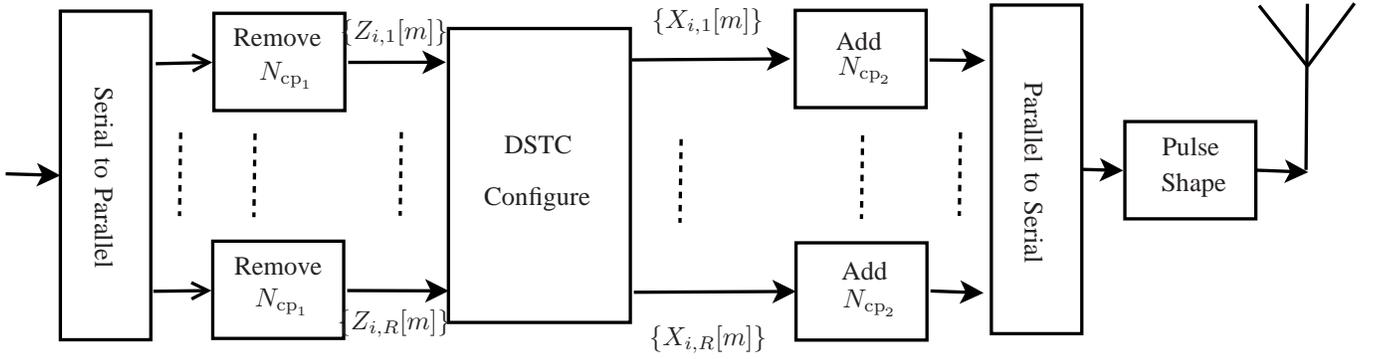},width=18cm}}
\caption{Configuration process at each relay, for $i=1,\cdots,R$ and $m=0,\cdots,N-1$.}
\label{fig:relaysblk}
\end{figure*}

Next, the results of the differential encoding block are configured into $R$ parallel sequences $\{s_1[n]\},\cdots,\{s_R[n]\},$ each with a length of $N$, i.e., $\quad n=0,\cdots,N-1$. Then, the N-point IDFT is applied to $\{s_r[n]\}, \quad r=1,\cdots,R$ to obtain the following OFDM symbols
\begin{equation}
\begin{split}
S_r[m]=\IDFT\{s_r[n]\}=\frac{1}{\sqrt{N}} \sum \limits_{n=0}^{N-1}s_r[n]\exp\left(j\frac{ 2\pi nm}{N}\right),
\end{split}
\end{equation}
for $r=1,\cdots, R$ and $m=0,\cdots,N-1$. Next, the last $N_{\cp_1}\geq (L-1)$ OFDM symbols in each sequence is appended to its beginning to obtain 
\begin{equation}
\label{eq:Srcp}
\{S_r[N-N_{\cp_1}],\cdots,S_r[N-1],S_r[0],\cdots,S_r[N-1]\},
\end{equation} 
for $r=1,\cdots,R$. The resulting $R$ parallel OFDM sequences of length $(N+N_{\cp_1})$ are then converted to a serial sequence of OFDM symbols. Finally, the discrete-time OFDM symbols are band-limited using a pulse-shaping filter,  multiplied by $\sqrt{P_0R}$ and passed to a radio-frequency (RF) block for up-conversion and transmission to the relays over $R(N+N_{\cp_1})T_s$ seconds, in Phase I. Here, $P_0$ is the average transmit power per OFDM symbol at the source and $T_s$ is the symbol duration. As mentioned before, the source has only one antenna.

\subsection{Relays' Operation}
The transmitted signals from the source go through frequency-selective fading channels and are received at the relays. The frequency-selective fading coefficients are assumed to be quasi-static during the transmission of $R$ sub-blocks or sequences, i.e., the coherence interval of $R(N+N_{\cp_1})T_s$ is required. 

The received signals at each relay are down-converted by an RF block, passed through the matched-filter and sampled at the symbol rate to obtain discrete-time samples. As the relays are individually synchronized to the source during Phase I, no ISI is caused by synchronization errors at the relays. The discrete-time samples are processed at each relay, independently, as illustrated in Fig.~\ref{fig:relaysblk}. As shown, first, the received serial sequence of length $R(N+N_{\cp_1})$ is converted to $R$ parallel sequences of length $(N+N_{\cp_1})$. Then, $N_{\cp_1}$ samples are removed from the beginning of each sequence. The result can be expressed as
\begin{equation}
\label{eq:Zij}
\begin{split}
Z_{i,r}[m]&=\sqrt{P_0R} \sum \limits_{l=0}^{L-1} q_{i,l} S_r[\modN{m-l}]+\Psi_{i,r}[m] \\
&=\sqrt{ P_0 R} \left(q_{i}[m] \otimes S_{r}[m]\right)+\Psi_{i,r}[m],\\
\end{split}
\end{equation}
for $i,r=1,\cdots, R$, and $m=0,\cdots, N-1$. Here, $\Psi_{i,r}[m] \sim \CN(0,N_0)$ is the noise element at the $i$th relay, sub-block $r$ and symbol-index $m$.

Let us denote $Z_{i,r}[\modN{-m}]$ as the \emph{circular time-reversal} of $Z_{i,r}[m]$ defined as \cite{DSP-Boaz}
\begin{equation}
\label{eq:CTR}
Z_{i,r}[\modN{-m}]= \left\lbrace 
\begin{matrix}
Z_{i,r}[0], & m=0 \\
Z_{i,r}[N-m], & \mbox{otherwise},
\end{matrix} \right.
\end{equation} 
for $i,r=1,\cdots,R$ and $m=0,\cdots,N-1$. Note that circular time-reversal is different from regular time-reversal \cite{DSP-Boaz}.

The samples at the $i$th relay are then configured as
\begin{equation}
\label{eq:Xij}
\begin{bmatrix}
X_{i,1}[m] \\
\vdots \\
X_{i,R}[m] \\
\end{bmatrix}
= A \left(  \B_i
\begin{bmatrix}
Z_{i,1}[m] \\
\vdots \\
Z_{i,R}[m]
\end{bmatrix}
+
\C_i
\begin{bmatrix}
Z_{i,r}^*[\modN{-m}] \\
\vdots \\
Z_{i,r}^*[\modN{-m}]
\end{bmatrix}
\right)
\end{equation}
for $m=0,\cdots,N-1$, where $A=\sqrt{P_{\rt}/(P_0+N_0)}$ is the amplification factor and $P_{\rt}$ is the transmit power per symbol at relays. Typically, a total power $P$ is divided between the source and relays as $P_0=P/2, P_{\rt}=P/(2R)$. Also, $\B_i$ and $\C_i$ are the dispersion matrices determined based on the unitary matrix that is used for the network. The choice of $\B_i$ and $\C_i$ is such that an effective space-time block code \cite{stc-HJ} is constructed at the destination. Interestingly, the dispersion matrices designed for conventional DSTC in \cite{D-DSTC-Y,DSTC-HJ,DSTC-Y} can be used in our system as well. For example for a network with two relays, $\B_i$ and $\C_i$ can be chosen as \cite{D-DSTC-Y}
\begin{equation}
\label{eq:Alam-B1C1}
\B_1=
\begin{bmatrix}
1 & 0 \\
0 & 1
\end{bmatrix},\;
\C_1=\0, \;
\B_2=\0,\;
\C_2=
\begin{bmatrix}
0 & -1 \\
1 & 0
\end{bmatrix},
\end{equation}
to construct a $2\times 2$ orthogonal code at the destination. Thus, for a network with two relays using \eqref{eq:Alam-B1C1}, the configurations at Relays 1 and 2 are
\begin{equation}
\label{eq:Xij-2Relays}
\begin{split}
&X_{1,1}[m]=A Z_{1,1}[m], \\
&X_{1,2}[m]=A Z_{1,2}[m], \\
&X_{2,1}[m]= -A {Z}_{2,2}^*[\modN{-m}], \\
&X_{2,2}[m]= A {Z}_{2,1}^*[\modN{-m}],
\end{split}
\end{equation}
for $m=0,\cdots,N-1$.

Next, the last $N_{\cp_2}$ samples of sequences $\{X_{i,r}[m]\}$,  are appended to their beginnings as the cyclic prefix to obtain 
$$
\{X_{i,r}[N-N_{\cp_2}],\cdots,X_{i,r}[N-1],X_{i,r}[0],\cdots,X_{i,r}[N-1]\}
$$ 
with length $N+N_{\cp_2}$ for $i,r=1,\cdots,R$. Typically, $N_{\cp_2}\geq (L-1)$ should be chosen to resolve the ISI caused by frequency-selective channels. However, this condition may be refined to overcome synchronization errors at the destination as will be discussed shortly. Finally, $R$ parallel sequences are converted to a serial sequence of length $R(N+N_{\cp_2})$ at each relay. The discrete-time samples are then band-limited using the pulse-shaping filter and passed to the RF block to be up-converted. In Phase II, the relays broadcast their signals simultaneously to the destination. As mentioned before, each relay is equipped with one antenna.

\subsection{Decoding Process at Destination}
\label{subsec:Dest}
\begin{figure*}[t!]
\psfrag {r11} [] [] [1] {$X_{1,r}[-1]$}
\psfrag {r12} [] [] [1] {$X_{1,r}[0]$}
\psfrag {r13} [] [] [1] {$X_{1,r}[1]$}
\psfrag {r14} [] [] [1] {$X_{1,r}[2]$}

\psfrag {r21} [] [] [1] {$X_{i,r}[-d_i-1]$}
\psfrag {r22} [] [] [1] {$X_{i,r}[-d_i]$}
\psfrag {r23} [] [] [1] {$X_{i,r}[-d_i+1]$}
\psfrag {r24} [] [] [1] {$X_{i,r}[-d_i+2]$}

\psfrag {r31} [] [] [1] {$X_{R,r}[-d_R-1]$}
\psfrag {r32} [] [] [1] {$X_{R,r}[-d_R]$}
\psfrag {r33} [] [] [1] {$X_{R,r}[-d_R+1]$}
\psfrag {r34} [] [] [1] {$X_{R,r}[-d_R+2]$}

\psfrag {T} [b] [] [1] {$T_s$}
\psfrag {tau1} [] [] [1] {$\tau_i$}
\psfrag {tau2} [] [] [1] {$\tau_R$}
\centerline{\epsfig{figure={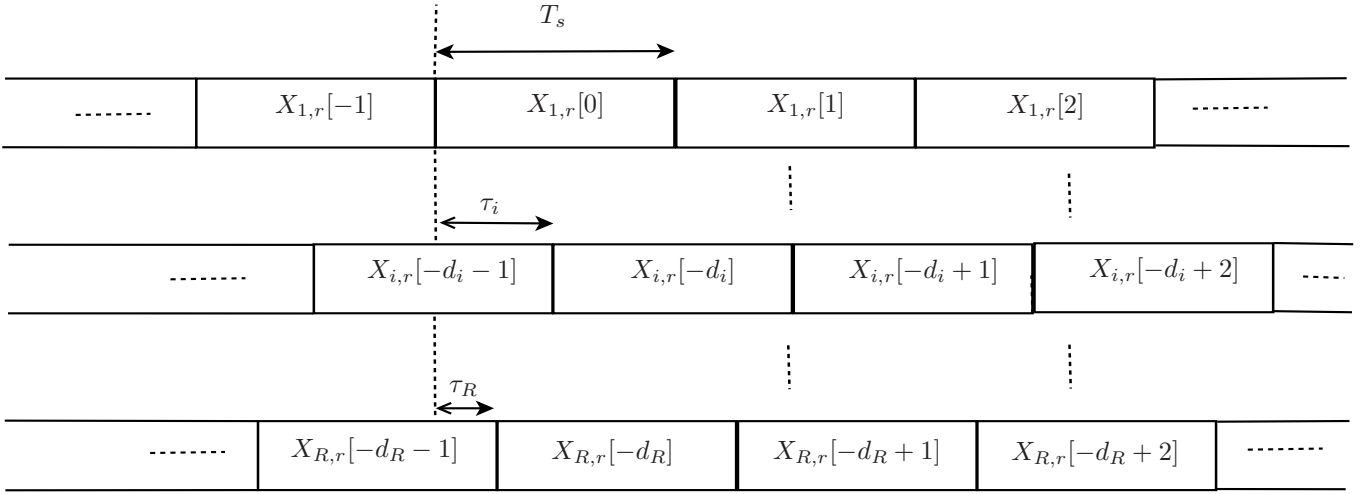},height=6.5cm,width=18cm}}
\caption{Received signals from the relays in the first path at the destination. The destination is synced with Relay 1. The received signal from Relay $i$ is $(d_iT_s+\tau_i)$ seconds late with respect to Relay 1, $i=2,\cdots,R$, $d_i\in \Zc$ and $0 \leq \tau_i\leq T_s$, $r=1,\cdots,R$.}
\label{fig:Rsigs}
\end{figure*}

Conventionally, it is assumed that the signals from the relays arrive at the same time at the destination. However, due to the distributed nature of relay networks, this assumption is easily violated. Here, without loss of generality, we assume that the destination is synced with the first relay. The signals from the $i$th relay, $i=2,\cdots,R$, may arrive $(d_iT_s+\tau_i)$ seconds late with respect to that of the firs relay, where $d_i$ is an integer number, $T_s$ is the symbol duration and $0\leq\tau_i < T_s$. This scenario is depicted for the first path, i.e. $l=0$, in Fig.~\ref{fig:Rsigs}. It is assumed that a similar scenario happens for the other multipaths, i.e., $l=1,\cdots,L-1$.

\begin{figure}[t]
\psfrag {x111111} [] [] [0.8] {$X_{1,r}[0]$}
\psfrag {x222222} [] [] [0.8] {$X_{1,r}[-1]$}
\psfrag {x333333} [] [] [0.8] {$X_{1,r}[1]$}

\psfrag {x444444444} [] [] [0.8] {$X_{2,r}[-d_2]$}
\psfrag {x555555555} [] [] [0.8] {$X_{2,r}[-d_2-1]$}
\psfrag {x666666666} [] [] [0.8] {$X_{2,r}[-d_2+1]$}
\psfrag {Time} [] [] [0.8] {$t/T_s$}
\centerline{\epsfig{figure={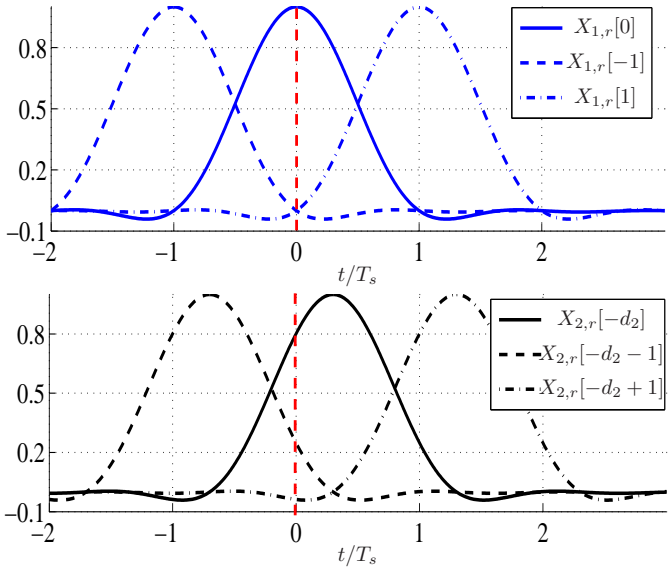},height=7.5cm,width=9cm}}
\caption{Received signals from Relay 1 (top) and Relay 2 (bottom) at the destination after the raised-cosine matched-filter with roll-off factor $\beta=0.9$, when $\tau_1=0, \tau_2=0.3T_s$, symbol rate sampling point is highlighted with the vertical dashed line.}
\label{fig:mf1}
\end{figure}

The received signals at the destination are passed through a matched-filter and sampled at the symbol rate. Depending on system specifications and requirements, different filters such as rectangular, sinc, raised-cosine or Gaussian can be used. For instance, consider a raised-cosine pulse-shape and matched-filter $p(t)=\sinc(t/T_s)\cos(\pi \beta t/T_s)/(1-4\beta^2t^2/T_s^2),$ \cite{madhow} with roll-off factor $\beta$ and $t$ as time. This is a practical filter as its time-domain side-lobes and frequency bandwidth can be limited by choosing an appropriate $\beta$. In Fig.~\ref{fig:mf1}, the base-band signals in the first path, assuming two relays and the raised-cosine pulse-shape and matched-filter with $\beta=0.9$ are depicted. Note that, for clear illustration, the two signals are plotted separately and the effect of multipaths is not shown in Fig.~\ref{fig:mf1}. The sampled signal after the matched-filter is the super-position of both signals. One signal from the first relay, whose peak value is at the sampling point, contributes to the sampled signal. The side-lobes of the first-relay's signal cross zero as $\tau_1=0$. Also, three fractions of the signal from the second relay contribute to the sampled signal. These signals are proportional to $p(-T_s-\tau_2)$, $p(-\tau_2)$ and $p(T_s-\tau_2)$. A similar figure can be plotted for other multipaths. Note that, depending on the filter and its side-lobes in the time-domain, more/less terms may appear in the sampled signal. For a general case, let us denote the number of significant side-lobes of the matched-filter with $L_{\mf}$. As an example, for the raised-cosine filter depicted in Fig.~\ref{fig:mf1}, $L_{\mf}=1$ as the contributions of the higher-order side-lobes of $p(t)$, beyond the first lobe, are very small and are thus ignored. 

Based on the above observations, it can be seen that in addition to the ISI caused by frequency-selectivity, another ISI appears due to symbols' misalignment. Therefore, to handle both ISIs, the cyclic prefix length at the relays should be chosen based on the multipath length ($L$), integer part of delays ($d_i$) and the number of significant side-lobes. In practice, based on the propagation environment, the maximum delay $d_{\max}=\max\limits_{i=1,\cdots,R} \{d_i \}$ is assumed to be known. Hence, the cyclic prefix length $N_{\cp_2}\geq (L-1+ d_{\max}+2L_{\mf})$ can be chosen at the relays.

\begin{figure}[t]
\psfrag {v1} [] [] [1] {$\qquad \{\widehat{v}_1[n]\}$}
\psfrag {v2} [] [] [1] {$\qquad \{\widehat{v}_R[n]\}$}
\psfrag {StP} [] [] [1][-90] {Serial to Parallel}
\psfrag {DD} [] [] [1][-90] {Differential Decoding}
\psfrag {DFT} [] [] [1] {DFT}
\psfrag {RCP} [] [] [1] {Remove}
\psfrag {Ncp2} [] [] [1] {$N_{\cp_2}$}
\psfrag {Y1} [b] [] [1] {$\quad\{Y_1[m]\}$}
\psfrag {Y2} [t] [] [1] {$\{Y_R[m]\}$}
\psfrag {y1} [b] [] [1] {$\quad\{y_1[n]\}$}
\psfrag {y2} [t] [] [1] {$\{y_R[n]\}$}
\centerline{\epsfig{figure={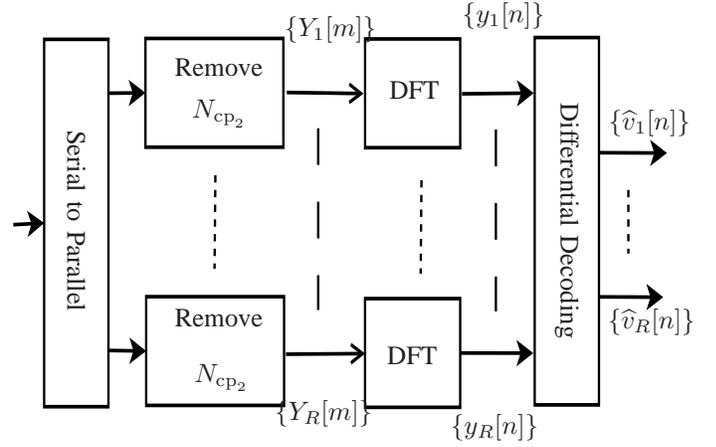},height=5.5cm,width=8.75cm}}
\caption{Decoding process at the destination, for $n,m=0,\cdots,N-1$.}
\label{fig:destblk}
\end{figure}

The sampled signal after the matched filter is used for decoding. The block diagram of the decoding process at the destination is depicted in Fig.~\ref{fig:destblk}. As seen, the received serial sequence of length $R(N+N_{\cp_2})$ is converted to $R$ parallel sequences of length $(N+N_{\cp_2})$. Next, the first $N_{\cp_2}$ samples of each sequence are removed. The remaining samples can be expressed as
\begin{equation}
\label{eq:Yr[m]}
\begin{split}
Y_r[m]&=\sum \limits_{i=1}^R \sum\limits_{\tl=-L_{\mf}}^{L_{\mf}} p(\tl T_s-\tau_i) \left( g_i[m] \otimes X_{i,r}[m-d_i-\tl]\right)
\\ &+ \Phi_r[m],
\end{split}
\end{equation}
for $ m=0,\cdots,N-1$ and $r=1,\cdots,R$, where $\Phi_r[m]\sim \CN(0,N_0)$. If $\tau_i=0$ for $i=1,\cdots,R$, the internal sum in \eqref{eq:Yr[m]} is reduced to one term. Otherwise, for instance when $L_{\mf}=1$, the effect of synchronization errors are observed in $p(-T_s-\tau_i)$, $p(-\tau_i)$ and $p(T_s-\tau_i)$ for $i=1,\cdots,R$. This means that the sampled signal after the match-filter is scaled proportional to the matched-filter response at the delay values.

Next, the DFT is applied to each sequence as
\begin{equation}
\label{eq:dftYm}
\begin{split}
y_r[n]=\DFT\{Y_r[m]\}=\frac{1}{\sqrt{N}} \sum \limits_{m=0}^{N-1}Y_r[m]\exp\left(-j\frac{ 2\pi mn}{N}\right),
\end{split}
\end{equation}
to obtain
\begin{equation}
\label{eq:y_r[n]}
\begin{split}
y_r[n]=& \sum \limits_{i=1}^{R}\sum\limits_{\tl=-L_{\mf}}^{L_{\mf}}  p(\tl T_s-\tau_i) G_i[n] \et^{-j\frac{2\pi n (d_i+\tl)}{N}} x_{i,r}[n]\\+&\phi_r[n] 
= \sum \limits_{i=1}^R \widetilde{G}_i[n] x_{i,r}[n]+\phi_r[n] 
\end{split}
\end{equation}
for $r=1,\cdots,R$, $n=0,\cdots,N-1$, where
\begin{equation}
\label{eq:GnPn}
\begin{split}
G_i[n]&=\sum \limits_{l=0}^{L-1} g_{i,l} \et^{-j \frac{2\pi nl}{N}},\\
\widetilde{G}_i[n]&= G_i[n] \et^{ -j\frac{2\pi n d_i}{N}} \sum\limits_{\tl=-L_{\mf}}^{L_{\mf}} p(\tl T_s-\tau_i) \et^{ -j\frac{2\pi n \tl}{N}}, \\
x_{i,r}[n]&=\DFT\{ X_{i,r}[m]\}, \; \phi_r[n]=\DFT\{\Phi_r[m]\}.
\end{split}
\end{equation}
Note that $\phi_r[n]\sim \CN(0,N_0)$.

In the matrix form, one can write
\begin{equation}
\label{eq:y[n]}
\y[n]=\left[ \; \x_1[n],\cdots,\x_R[n]\; \right] \widetilde{\G}[n]+\phib[n],
\end{equation}
where 
\begin{equation}
\label{eq:yxG}
\begin{split}
\y[n]&= [ \; y_1[n],\cdots, y_R[n]\;]^t, \\
\x_i[n]&= [ \; x_{i,1}[n],\cdots, x_{i,R}[n]\;]^t, \\
\widetilde{\G}[n]&=[ \; \tG_{1}[n],\cdots, \tG_{R}[n]\;]^t, \\
\phib[n]&=[ \; \phi_{1}[n],\cdots, \phi_{R}[n]\;]^t. 
\end{split}
\end{equation}

After some manipulation (see details in the Appendix), \eqref{eq:y[n]} can be shown to be 
\begin{equation}
\label{eq:y[n]-2}
\y[n]=A \sqrt{P_0R} \Sb[n] \Hb[n] + \w[n],
\end{equation}
where 
\begin{equation}
\label{eq:HSw}
\begin{split}
\Sb[n]&=\left[\; \widehat{\B}_1 \widehat{\s}_1[n], \cdots, \widehat{\B}_R \widehat{\s}_R[n]   \right], \\
\Hb[n]&=\left[\; H_1[n],\cdots, H_R[n]\;\right]^t, \\
H_i[n]&=\widehat{Q}_i[n] \widetilde{G}_i[n],\\
\w[n]&=\sum \limits_{i=1}^R \widetilde{G}_i[n] \widehat{\B}_i \widehat{\psib}_i[n]+\phib[n],
\end{split}
\end{equation}
with
\begin{equation}
\left.
\begin{matrix}
\widehat{\B}_i=\B_i \\
\widehat{\s}_i[n]=\s[n] \\
\widehat{Q}_{i}[n]=Q_{i}[n] \\
\widehat{\psib}_i[n]=\psib_i[n]
\end{matrix}
\right\rbrace
\mbox{\;if} \; \C_i=\0,
\end{equation}
\begin{equation}
\left.
\begin{matrix}
\widehat{\B}_i=\C_i \\
\widehat{\s}_i[n]=\s^*[n] \\
\widehat{Q}_{i}[n]=Q_{i}^*[n] \\
\widehat{\psib}_i[n]=\psib_i^*[n]
\end{matrix}
\right\rbrace
\mbox{\;if} \; \B_i=\0, 
\end{equation}
and 
\begin{equation}
\begin{split}
\s[n]&=[\; s_1[n], \cdots, s_R[n] \;]^t,\\
Q_i[n]&=\sum \limits_{l=0}^{L-1} q_{i,l} \exp(-j2\pi nl/N), \\
\psib_i[n]&=[\; \psi_{i,1}[n], \cdots, \psi_{i,R}[n] \;]^t, \\
\psi_{i,r}[n]&=\DFT\{\Psi_{i,r}[m]\},
\end{split}
\end{equation}
for $i,r=1,\cdots,R$ and $0 \leq n \leq N-1$. $\Hb[n]$ and $\w[n]$ are the equivalent channel and noise, at sub-carrier $n$, respectively. It is noted that $\psi_{i,r}[n]\sim \CN(0,N_0)$. Thus, the ISI effect caused by frequency-selectivity and synchronization errors is completely removed.

For instance, for a network with two relays, \eqref{eq:y[n]-2} can be simplified as
\begin{equation}
\label{eq:yn-2R}
\begin{bmatrix}
y_1[n] \\ y_2[n]
\end{bmatrix}
=A \sqrt{2P_0} \begin{bmatrix}
s_{1}[n] & -s_{2}^*[n] \\
s_{2}[n] & s_{1}^*[n]
\end{bmatrix}
\begin{bmatrix}
H_1[n] \\
H_2[n]
\end{bmatrix}+
\begin{bmatrix}
w_1[n] \\
w_2[n]
\end{bmatrix},
\end{equation}
with
\begin{equation}
\label{eq:H1w1w2}
\begin{split}
H_1[n]&=Q_1[n] \tG_1[n],\; H_2[n]=Q_2^*[n] \tG_2[n],\\
w_1[n]&=A \left( \tG_1[n] \psi_{1,1}[n]- \widetilde{G}_{2}[n] \psi_{2,2}^*[n]  \right)+\phi_1[n],\\
w_2[n]&=A \left( \tG_1[n] \psi_{1,2}[n]+ \widetilde{G}_{2}[n] \psi_{2,1}^*[n]  \right)+\phi_2[n].
\end{split}
\end{equation}
for $0 \leq  n \leq N-1$. 

\textbf{Proposition 1.} The above D-OFDM DSTC system achieves a general diversity of $(R,-R)$.
\begin{proof}
For each sub-carrier, the system equation of \eqref{eq:y[n]-2} is similar to that of the conventional DSTC \cite{DSTC-Y,D-DSTC-Y,DSTC-HJ} over flat-fading channels with perfect synchronization. Also, $\Sb[n]$ is a space-time block code which is constructed by the distributed relays at the destination at sub-carrier $n$. In other words, the proposed system over frequency-selective channels becomes equivalent to $N$ parallel conventional D-DSTC over flat-fading channels with perfect synchronization. Therefore, by using arguments in \cite{DSTC-Y,DSTC-HJ,D-DSTC-Y}, the pair-wise error probability (PEP) can be written as
\begin{equation}
\label{eq:pep}
\PEP \propto P^{-R\left(1-\frac{\log (\log (P))}{\log (P)} \right)}.
\end{equation}
Using the formulation introduced in \cite{erdem-DB} for error rates with logarithmic terms, the first-order diversity, i.e., the conventional diversity is
\begin{equation}
\label{eq:d1}
d_1= \lim \limits_{P \rightarrow \infty}  -\frac{\log (\PEP)}{\log (P)}=R.
\end{equation}
Also, the second-order diversity is 
\begin{equation}
\label{eq:d2}
d_2= \lim \limits_{P \rightarrow \infty} -\frac{\log(\PEP)+d_1\log(P)}{\log(\log(P))}= -R.
\end{equation}
The achieved diversity gain is defined as $d=(d_1,d_2)=(R,-R)$ \cite{erdem-DB} and the asymptotic performance, at high transmit power, can be expressed as
\begin{equation}
\label{eq:PEP-d1d2}
\PEP \propto (\log(P))^{R} P^{-R}.
\end{equation}
\end{proof}

For given $\{g_{i,l}\},\; i=1,\cdots,R$, the equivalent noise $\w[n] \sim \CN(\0, \sigma^2[n] \I_R)$, where 
\begin{align}
\label{eq:cn}
\sigma^2[n]&=N_0\left(1+A^2\sum \limits_{i=1}^R |\tG_i[n]|^2\right).
\end{align}
Also, the received SNR per symbol, for given $\{g_{i,l}\}$ can be obtained as 
\begin{equation}
\label{eq:gama}
\gamma\left(n,\{\tau_i\}_{i=1}^R\right)= A^2P_0 \frac{ \sum \limits_{i=1}^R |\tG_i[n]|^2}{\sigma^2[n]},
\end{equation}
for $n=0,\cdots,N-1$. This shows that the effective SNR is a function of both sub-carrier number and delay values $\tau_i,\; i=1,\cdots,R$ and independent of $d_i$. 

\begin{figure}[b]
\psfrag {n} [] [] [1] {$n$}
\psfrag {tau00} [l] [] [1] {$\tau_2=0\&T_s$}
\psfrag {tau02} [l] [] [1] {$\tau_2=(0.2\&0.8)T_s$}
\psfrag {tau03} [l] [] [1] {$\tau_2=(0.3\&0.7)T_s$}
\psfrag {tau04} [l] [] [1] {$\tau_2=(0.4\&0.6)T_s$}
\psfrag {tau05} [l] [] [1] {$\tau_2=0.5T_s$}
\psfrag {gama} [] [] [1] {$\gamma(n,\tau_2)$, dB}
\centerline{\epsfig{figure={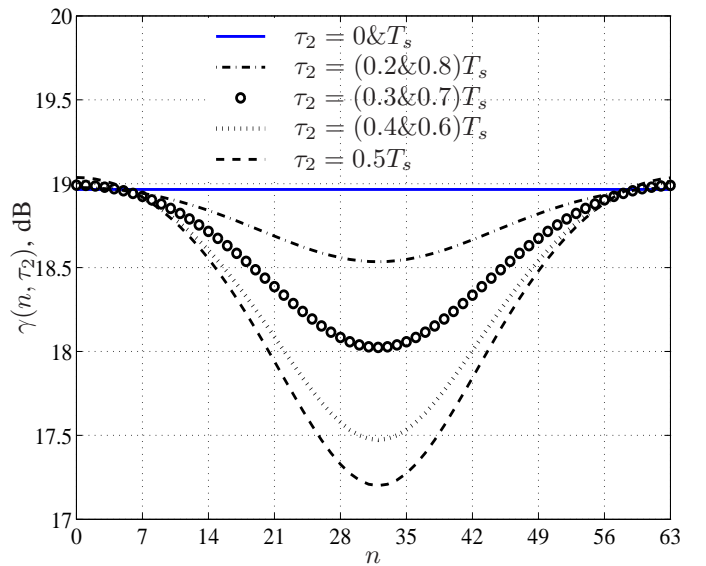},height=7.5cm,width=9cm}}
\caption{Average received SNR vs. $n$ and $\tau_2$ for a network with two relays over flat-fading channels, when $N=64$, $P/N_0=25$dB, $P_0=P/2,P_\rt=P/4$, $|g_{1,0}|^2=|g_{2,0}|^2=1$.}
\label{fig:fig:gama}
\end{figure}

If all channels are flat-fading, i.e., $L=1$ and all relays are perfectly synchronized at the symbol level, i.e., $\tau_i=0,\; i=1,\cdots,R$, with the raised-cosine filter, $p(\tau_i)=1$, $p(T_s-\tau_i)=0$ and we will have $\tG_i[n]=g_i[n]=g_{i,0}$. In this case, the noise variance and the received SNR of the proposed system are the same as that of the conventional D-DSTC \cite{D-DSTC-Y} for $\tau_i=0$ over flat-fading channels. However, for $\tau_i \neq 0$ the average received SNR is a function of $\tau_i$ and $n$. To see this dependency, the received SNR, i.e. \eqref{eq:gama}, is plotted versus $n$ in Fig.~\ref{fig:fig:gama} for a network with two relays $(R=2)$ over flat-fading channels. It is assumed that $\tau_1=0, \tau_2=\{0,0.2,0.4,0.6,0.8,1\}$, $N=64$, $P/N_0=25$ dB, $P_0=P/2,P_\rt=P/4$ and for simplicity $|g_{1,0}|^2=|g_{2,0}|^2=1$. As can be seen from the figure, SNR decreases with $n$ and reaches a minimum at $n=N/2$. Also, overall, SNR decreases with increasing $\tau_2$ and reaches its minimum value at $\tau_2=0.5T_s$. Then it increases with increasing $\tau_2$ towards $T_s$ such that $\gamma(n,\tau_2)=\gamma(n,T_s-\tau_2)$. This phenomena yields the same average BER for symmetric values of $\tau_2$ around $0.5 T_s$, as will be seen in the simulation results. It should be mentioned that the performance is independent of $d_i$ values as long as the cyclic-prefix length is long enough. 

By writing \eqref{eq:y[n]-2} for two consecutive block-indexes $(k),(k-1)$, using $\Sb[n]^{(k)}=\Vb[n]^{(k)} \Sb[n]^{(k-1)}$ and assuming that $\Hb[n]$ is constant over two consecutive blocks, 
one obtains
\begin{equation}
\label{eq:ykykm1}
\y[n]^{(k)}=\Vb[n]^{(k)} \y[n]^{(k-1)}+\widetilde{\w}[n]^{(k)},
\end{equation}
where 
\begin{equation}
\label{eq:wtn}
\widetilde{\w}[n]^{(k)}=\w[n]^{(k)}-\Vb[n]^{(k)} \w[n]^{(k-1)},
\end{equation}
for $n=0,\cdots,N-1$. It is easy to see that, for given $\{g_{i,l}\},\; i=1,\cdots,R$, the overall noise $\widetilde{\w}[n]^{(k)}$ is  $\CN(\0,2\sigma^2[n]\I_R)$.  

Therefore, the non-coherent decoding can be applied as
\begin{equation}
\label{eq:decod-ofdm}
\widehat{\Vb}[n]=\arg \min \limits_{\Cbc} \|\y[n]^{(k)}-\Vb[n]^{(k)} \y[n]^{(k-1)}\|,
\end{equation}
for $ n=0,\cdots,N-1$, to decode the $N$ codewords or equivalently $RN$ data symbols without any channel information or knowing the amount of delays. 

Note that the decoding complexity of the proposed method is the same as that of the conventional D-DSTC \cite{D-DSTC-Y}. For orthogonal designs Eq. \eqref{eq:decod-ofdm} can be simplified to obtain a symbol-by-symbol decoding. For instance, for $2\times2$ orthogonal designs, using Eq. \eqref{eq:yn-2R} and Eq. \eqref{eq:decod-ofdm}, by computing decision metrics as
\begin{equation}
\label{eq:sym-decod1}
\begin{split}
\tilde{v}_1[n]&= y_1[n]^{(k)} y_1^*[n]^{(k-1)}+y_2^*[n]^{(k)} y_2[n]^{(k-1)},\\
\tilde{v}_2[n]&= y_2[n]^{(k)} y_1^*[n]^{(k-1)}-y_1^*[n]^{(k)} y_2[n]^{(k-1)},
\end{split}
\end{equation}
for $n=0,\cdots,N-1$, fast symbol-by-symbol non-coherent decoding can be obtained.

It is pointed out that, approximately a coherence interval of $3RN$ symbols is required for non-coherent decoding. Here, $N\gg \max\{N_{\cp_1},N_{\cp_2}\}$ to reduce the data loss of cyclic-prefix. Hence, a trade-off between the coherence time and data rate should be used to choose $N$.

In the system developed so far, the output of the matched filter is sampled at the symbol rate. {For future reference, this system is referred to as D-OFDM1 DSTC.} In the next section, we consider a new sampling scheme to improve the system performance.

\subsection{Double Sampling Scheme}
Our proposed method removes the ISI caused by frequency selectivity and both integer multiple and fractional synchronization errors. However, dependency of the received SNR to the fractional timing errors is inevitable. This dependency, as will be seen in the simulation results, leads to some performance loss compared with the perfect synchronization case. To compensate this gap, we propose to sample the received signal after the matched-filter as depicted in Fig.~\ref{fig:new_sample}. In other words, in addition to the conventional sampling at $0,\pm T_s,\pm 2T_s,\cdots$, additional samples are taken at $\pm 1/2T_s,\pm 3/2T_s,\pm 5/2T_s,\cdots$. {Similar to \eqref{eq:Yr[m]}, the additional sample can be expressed as
\begin{multline}
\label{eq:Yr[m]2}
\breve{Y}_r[m]= \breve{\Phi}_r[m]+\\
\sum \limits_{i=1}^R \sum\limits_{\tl=-L_{\mf}}^{L_{\mf}}  p((\tl+0.5) T_s-\tau_i) \left( g_i[m] \otimes X_{i,r}[m-d_i-\tl]\right),
\end{multline}
for $r=1,\cdots,R$ and $m=0,\cdots,N-1$, where $\breve{\Phi}_r[m]\sim \CN(0,N_0)$. Maximum-ratio combining of the two samples requires knowledge of the synchronization errors. Since the synchronization errors are not known, an equal-gain combining method is employed to obtain
\begin{equation}
\label{eq:Yr[m]-new}
\begin{split}
Y_r[m]&+\breve{Y}_r[m]=\\ &
 \sum \limits_{i=1}^R \sum\limits_{\tl=-L_{\mf}}^{L_{\mf}}  \left(p(\tl T_s-\tau_i)+p((\tl+0.5) T_s-\tau_i) \right)\\ & \left( g_i[m] \otimes X_{i,r}[m-d_i-\tl]\right)+\Phi_r[m]+\breve{\Phi}_r[m],
\end{split}
\end{equation}
for $m=0,\cdots,N-1$, $r=1,\cdots,R$. The result is then used for decoding (input to Fig.~\ref{fig:destblk}). The rest of the decoding process at the destination remains the same. This only changes the definition of $\widetilde{G}_i[n]$ in \eqref{eq:GnPn} to
\begin{equation}
\label{eq:GnPn-new}
\begin{split}
\widetilde{G}_i[n]&=G_i[n] \exp\left( -j\frac{2\pi n d_i}{N}\right) \\  &\sum\limits_{\tl=-L_{\mf}}^{L_{\mf}} \left(p(\tl T_s-\tau_i)+p((\tl+0.5) T_s-\tau_i) \right) \et^{ -j\frac{2\pi n \tl}{N}},
\end{split}
\end{equation}
and $\phi_{r}[n]\sim \CN(0,2N_0)$. Also,
$
\sigma^2[n]=N_0\left(2+A^2\sum \limits_{i=1}^R |\tG_i[n]|^2\right).
$
The other equations remain the same.

Using double sampling and equal-gain combining, the absolute value of $\widetilde{G}_i[n]$ and hence the average received SNR are increased. Thus, the system becomes less susceptible to fractional synchronization errors. Note that the average received SNR, and thus the system performance, of the double sampling scheme at $\tau_i=0.5T_s$ is the same as that at $\tau_i=0,T_s$. As will be seen in the simulation results, the double sampling method overcomes the performance gap between perfect and imperfect synchronizations. The sampling process in Fig.~\ref{fig:new_sample} can be equivalently implemented by doubling the sampling rate. For future reference, this system is referred to as D-OFDM2 DSTC. Note that the decoder does not need to know the amount of synchronization errors to implement the double sampling scheme. 

\begin{figure}[t]
\psfrag {MF} [] [] [1] {Matched Filter}
\psfrag {RX} [] [] [1] {}
\psfrag {Ts} [] [] [1] {$T_s$}
\psfrag {D} [] [] [1] {$T_s/2$}
\centerline{\epsfig{figure={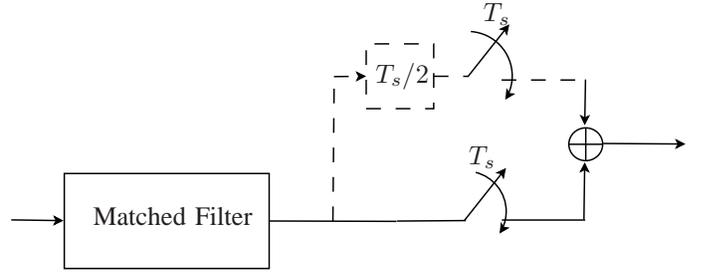},width=9cm}}
\caption{Double sampling scheme at the destination to improve the performance against fractional timing errors.}
\label{fig:new_sample}
\end{figure}

\subsection{Exploiting Intrinsic Diversity}
Frequency diversity can be attained by integrating channel coding and interleaving in the system. Here, we consider an $(r,1)$ repetition code, i.e., each bit of information is repeated $r$ times before being converted to modulation symbols \cite{gold_wireless}. It is important to interleave the coded bits so that information bits experience independent fading channels. A block interleaver is an array with $D_{\intv}$ rows and $r$ columns \cite{gold_wireless}. To satisfy the independent fading condition, $D_{\intv}>1/(f_DT_s)$ \cite{gold_wireless}, where $f_DT_s$ is the normalized Doppler frequency of the channels. Coded bits are read into rows of $r$ bits. The contents of the interleaver's columns are read out for modulation. The subsequent processing at the source and the relays is the same as that of the uncoded system. At the destination, the decoded symbols (output of Fig.~\ref{fig:destblk}) are read into the deinterleaver's columns. The deinterleaver is an array identical to the interleaver \cite{gold_wireless}. If soft decision metrics are available, such as in \eqref{eq:sym-decod1}, they can be directly read into the deinterleaver. Soft-decision decoding attains better performance than hard-decision decoding \cite{gold_wireless}. Next, the deinterleaver output is read out from its rows. Each row has $r$ decision metrics that can be combined using the maximum-ratio combining method. The output of the combiner is used for bit decoding. Consequently, similar to time diversity \cite{gold_wireless}, an $r$-fold frequency diversity gain is achieved. In other words, using an $(r,1)$ repetition code and an interleaver,  the achieved first-order diversity of D-OFDM DSTC is $rR$.

\section{Simulation Results}
\label{sec:Sim}
In this section, networks with $R=2,3,4$ relays are simulated in various scenarios including flat-fading and frequency-selective channels with synchronization errors. Through these simulations, the effectiveness of D-OFDM1 DSTC (symbol-rate sampling) and D-OFDM2 DSTC (double sampling) against ISI is illustrated. In the case of flat-fading channels, the results are compared with conventional DSTC \cite{D-DSTC-Y,DSTC-HJ} which have been considered for flat-fading channels with perfect synchronization. Moreover, for networks with $R=2$ relays over frequency-selective channels, channel coding and interleaving is integrated into the system to exploit additional frequency diversity.

First, networks with $R=2$ and $R=4$ relays are considered over flat-fading channels with synchronization errors. For the network with two relays, information bits are converted to BPSK symbols. Then, at the source, $2N$ symbols are encoded to $N$ unitary matrices as in \eqref{eq:alamouti}. The dispersion matrices as in \eqref{eq:Alam-B1C1} are used at the relays. For the network with four relays, the quasi-orthogonal design (QOD) defined as \cite{DSTC-HJ}
\begin{equation}
\label{eq:QOSTC}
\Vb[n]=\frac{1}{\sqrt{\sum \limits_{i=1}^{4} |v_i[n]|^2} } 
\begin{bmatrix}
v_1[n] & -v_2^*[n] & -v_3^*[n] & v_4[n] \\
v_2[n] &  v_1^*[n] & -v_4^*[n] & -v_3[n] \\
v_3[n] &  -v_4^*[n] & v_1^*[n] & -v_2[n] \\
v_4[n] &  v_3^*[n] &  v_2^*[n] & v_1[n] \\
\end{bmatrix}
\end{equation}
is employed to encode $4N$ data symbols for $n=0,\cdots,N-1$. The symbols $v_1[n],v_2[n]$ are picked from BPSK while $v_3[n],v_4[n]$ belong to the $\pi/2$-rotated BPSK constellation \cite{DSTC-HJ}. Note that in this configuration, the matrix defined in \eqref{eq:QOSTC} is unitary. Also, the dispersion matrices at the relays are determined as \cite{DSTC-HJ}
\begin{equation}
\label{eq:BiCi-R4}
\begin{split}
&\B_1=\I_4,\; \C_1=\0,
\\
&\B_2=\0,\; \C_2=
\begin{bmatrix}
0 & -1 & 0 & 0 \\
1 &  0 & 0 & 0 \\
0 &  0 & 0 & -1 \\
0 &  0 & 1 & 0 \\
\end{bmatrix},
\\
&\B_3=\0,\; \C_3=
\begin{bmatrix}
0 & 0 & -1 & 0 \\
0 &  0 & 0 & -1 \\
1 &  0 & 0 & 0 \\
0 &  1 & 0 & 0 \\
\end{bmatrix},
\\
&\B_4=
\begin{bmatrix}
0 & 0 & 0 & 1 \\
0 &  0 & -1 & 0 \\
0 &  -1 & 0 & 0 \\
1 &  0 & 1 & 0 \\
\end{bmatrix},\; \C_4=\0.
\end{split}
\end{equation}
Moreover, $N=64$ IDFT/DFT are employed. $N_{\cp_1}$ can be set to zero for flat-fading channels and $N_{\cp_2}$ can be chosen based on the synchronization errors and matched-filter. The integer parts of the synchronization errors ($d_i$) are randomly chosen based on the uniform distribution between one to five symbols, then $d_{\max}=5$. Also, the raised-cosine matched-filter with $\beta=0.9$ and $L_{\mf}=1$ is considered. To make the system symmetrical, $N_{\cp}=N_{\cp_1}=N_{\cp_2}= 7$ is chosen. Next, differential encoding is applied as \eqref{eq:sk-ofdm} to obtain two/four sequences of length $N$. Then, the $N$-point IDFT is applied to the two/four sequences of length $N$. Finally, $N_{\cp_1}$ symbols are appended. 

For flat-fading ($L=1$), the channel coefficients $q_{i,0},g_{i,0}\sim \CN(0,1),\; i=1,\cdots,R$, slowly change according to the Jakes' model with the normalized Doppler frequency of $f_DT_s=10^{-4}$. For the IEEE $802.11n$ standard, with carrier frequency of $f_c=2.45$GHz and symbol duration $T_s=4\mu$s, this amount of Doppler shift is equivalent to a velocity of $2.5$m/s (9km/hr). The simulation method of \cite{ch-sim} is used to generate the channel coefficients. The system is simulated for various amounts of timing errors $\tau_i=\tau$ for $i=2,\cdots,R$ where $\tau \in \{(0,0.2,0.3,0.4,0.5,0.6,0.7,0.8,1)T_s\}$ and $\tau_1=d_1=0$. At the destination, after the matched filter, both sampling at the symbol rate and also the double sampling scheme are applied.

Figs.~\ref{fig:M2_FF_R2} and \ref{fig:M2_FF_R4} depict the BER results of the D-OFDM1 and D-OFDM2 systems versus $P/N_0$ over flat-fading channels, for $R=2$ and $R=4$, respectively. For comparison purposes, the BER results of conventional D-DSTC \cite{D-DSTC-Y} ($d_i=0$ and various values of $\tau$) and coherent DSTC \cite{DSTC-HJ} (perfect synchronization $d_i=\tau=0$) with $R=2$ relays are plotted as benchmarks in Fig.~\ref{fig:M2_FF_R2}. Also, the BER curve of conventional D-DSTC \cite{D-DSTC-Y} (perfect synchronization) with $R=4$ relays is added to Fig.~\ref{fig:M2_FF_R4}.

\begin{figure}[t]
\psfrag {tau05} [l] [] [.8] {D-OFDM1, $ \tau=0.5T_s$}
\psfrag {tau04} [l] [] [.8] {D-OFDM1, $ \tau=(0.4 \& 0.6)T_s$}
\psfrag {tau03} [l] [] [.8] {D-OFDM1, $\tau=(0.3 \& 0.7)T_s$}
\psfrag {tau02} [l] [] [.8] {D-OFDM1, $\tau=(0.2 \& 0.8)T_s$}
\psfrag {tau00} [l] [] [.8] {D-OFDM1, $\tau=0 \& T_s$}
\psfrag {DDSTCt00} [] [r] [.8] { D-DSTC, $\tau=0$ }
\psfrag {DDSTCt02} [] [r] [.8] { $\qquad$ D-DSTC, $\tau=0.2T_s$ }
\psfrag {DDSTCt04} [] [r] [.8] { $\qquad$ D-DSTC, $\tau=0.4T_s$ }
\psfrag {DDSTCt06} [] [r] [.8] { $\qquad$ D-DSTC, $\tau=0.6T_s$ }
\psfrag {Coherent} [] [r] [.8] { $\qquad$ Coherent $\tau=0$ }
\psfrag {BER} [] [] [1] {BER}
\psfrag {Total Power} [] [] [1] {$P/N_0$ (dB)}
\psfrag {double05} [l] [] [.8] {D-OFDM2, $\tau=(0\&0.5\&1)T_s$}
\psfrag {double025} [l] [] [.8] {D-OFDM2, $\tau=0.25T_s$}
\centerline{\epsfig{figure={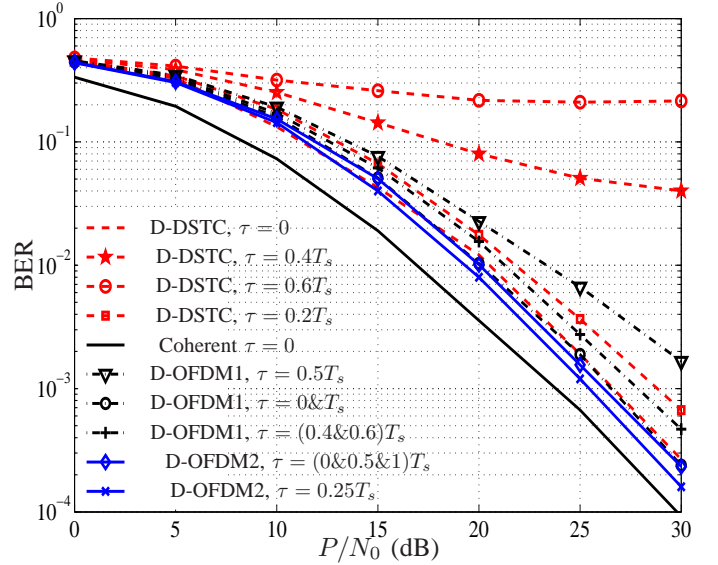},height=7.5cm,width=9cm}}
\caption{Simulation BER of a network with $R=2$ relays over flat-fading channels, D-OFDM1 (symbol-rate sampling), D-OFDM2 (double sampling) ($N=64,N_{\cp}=7$), D-DSTC \cite{D-DSTC-Y}, and coherent DSTC\cite{DSTC-HJ}, using \eqref{eq:alamouti} and BPSK, $d_1=\tau_1=0,\tau_2=\tau$, $d_2$ is uniformly random.}
\label{fig:M2_FF_R2}
\end{figure}

\begin{figure}[t]
\psfrag {double05} [l] [] [.8] {D-OFDM2, $\tau=(0\&0.5\&1)T_s$}
\psfrag {double025} [l] [] [.8] {D-OFDM2, $\tau=0.25T_s$}
\psfrag {tau05} [l] [] [0.8] {D-OFDM1, $\tau=0.5T_s$}
\psfrag {tau04} [l] [] [0.8] {D-OFDM1, $\tau=(0.4 \& 0.6)T_s$}
\psfrag {tau03} [l] [] [0.8] {D-OFDM1, $\tau=(0.3 \& 0.7)T_s$}
\psfrag {tau00} [l] [] [0.8] {D-OFDM1, $\tau=0 \& T_s$}
\psfrag {CDSTC} [l] [] [0.8] {D-DSTC, $\tau=0$ }
\psfrag {BER} [] [] [1] {BER}
\psfrag {Total Power} [] [] [1] {$P/N_0$ (dB)}
\centerline{\epsfig{figure={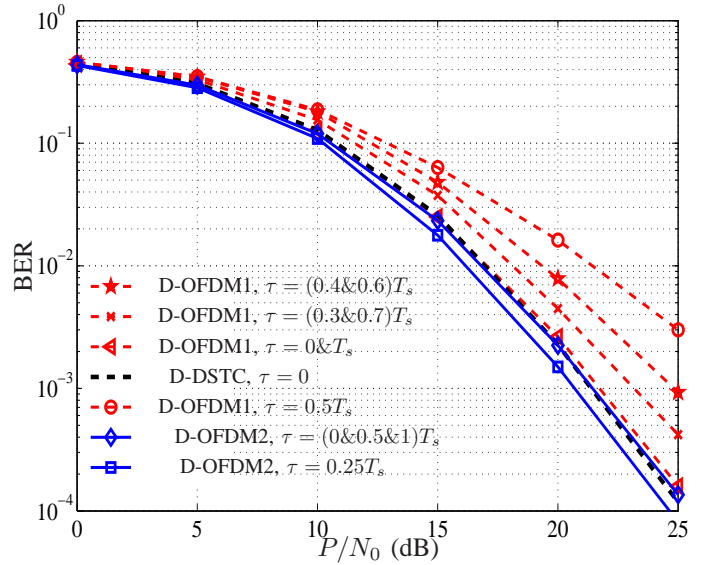},height=7.5cm,width=9cm}}
\caption{Simulation BER of a relay network with $R=4$ relays over flat-fading channels, D-OFDM1 (symbol-rate sampling), D-OFDM2 (double sampling) ($N=64, N_{\cp}=2$) and D-DSTC \cite{D-DSTC-Y}, using QOSTBC, $d_1=\tau_1=0$, $d_i$ is uniformly random, $\tau_i=\tau$ for $i=2,3,4$.}
\label{fig:M2_FF_R4}
\end{figure}

As seen in Fig.~\ref{fig:M2_FF_R2}, the performance of conventional D-DSTC \cite{D-DSTC-Y} with $R=2$ relays is severely degraded for $\tau>0.2T_s$ and an error floor appears in the BER curves. Obviously, if the synchronization error is greater than one symbol, i.e., $d_i>0$, the conventional D-DSTC will be totally ineffective. However, as shown in Figs.~\ref{fig:M2_FF_R2} and \ref{fig:M2_FF_R4}, D-OFDM1 DSTC is fairly robust against all values of the delays. The achieved diversity approaches the number of relays at high transmit power. As explained in Section~\ref{subsec:Dest}, the average received SNR of D-OFDM1 DSTC decreases with $\tau$ and reaches its minimum at $\tau=0.5T_s$. Hence, in the figures, the BER curves of D-OFDM1 DSTC gradually deviate from the benchmark. The largest gap corresponds to $\tau=0.5T_s$. Note that the BER curves of D-OFDM1 DSTC when $\tau=0$ and $T_s$ are similar to that of the conventional D-DSTC with perfect synchronization, although there is a slight rate loss due to the inclusion of the cyclic prefix.

The BER results of D-OFDM2 DSTC at $\tau=0.5T_s$ are close to that of the perfect synchronization. The system performance of D-OFDM2 DSTC is improved by about 6 dB in the worst case scenario ($\tau=0.5T_s$) with respect to that of D-OFDM1 DSTC. Also, the BER results of D-OFDM2 DSTC at $\tau =0.25T_s$ are slightly better than those at $\tau=0.5T_s$. In other words, D-OFDM2 DSTC is almost insusceptible to fractional synchronization errors. The performance difference between coherent DSTC \cite{DSTC-HJ} with perfect synchronization and that of D-OFDM DSTC in Fig.~\ref{fig:M2_FF_R2} is about 3 dB (the best one can expect).

Next, networks with $R=2$ and $R=4$ relays are considered over frequency-selective channels. For the network with $R=2$ relays, information bits are converted to BPSK symbols. BPSK and rotated BPSK constellations are used for the network with $R=4$ relays. Frequency-selective channels of length $L=6$ are assumed for both SR and RD links. The variance of the $L$ taps are equally normalized, i.e., $q_{i,l},g_{i,l}, \sim \CN(0,1/L)$ for $i=1,\cdots,R$ and $l=0,\cdots,L-1$. Also, $N=64$ and $N_{\cp_1}=N_{\cp_2}=12$ are chosen. The simulation results are plotted in Figs.~\ref{fig:M2_FS_R2} and \ref{fig:M2_FS_R4} for two and four relays, respectively. 

As can be seen in Figs.~\ref{fig:M2_FS_R2} and \ref{fig:M2_FS_R4}, the diversity approaches the number of relays at high transmit power. The effect of ISI caused by frequency-selective channels and synchronization errors is nullified. Again, some performance is lost due to the SNR reduction with $\tau$ in D-OFDM1 DSTC. This loss is totally compensated for in D-OFDM2 DSTC. Note that D-OFDM DSTC is not sensitive to the integer part $(d_i)$ of the synchronization errors. It should be mentioned that the plots of Figs.~\ref{fig:M2_FS_R2} and \ref{fig:M2_FS_R4} correspond to an uncoded system. Also, the variances of the $L$ taps in the frequency-selective channels are normalized to one. As such, the achieved diversity of the uncoded system over both flat-fading and frequency-selective channels is the same. However, the inherent frequency diversity can be exploited using channel coding and interleaving as will be seen shortly.

In addition, we consider a network with $R=3$ relays over flat-fading channels. In such a network, circulant codes \cite{D-DSTC-Y} can be employed. The data and dispersion matrices can be obtained from \cite[Sec.II-D]{D-DSTC-Y} and are not repeated here for reasons of space. We used the rotated-BPSK with the optimal angles listed in \cite [Table I]{D-DSTC-Y}. The simulation results are depicted in Fig.~\ref{fig:M2_FS_R3}. The block-error rate (BLER) of conventional D-DSTC \cite[Fig.4]{D-DSTC-Y}, using circulant codes with ideal synchronization over flat-fading channels is also plotted in Fig.~\ref{fig:M2_FS_R3}. It can be seen that, as expected, the diversity approaches three at high transmit power. The effect of ISI is nullified. The results of D-OFDM1 DSTC are the same as those of conventional D-DSTC \cite{D-DSTC-Y} at $\tau=0$. This means that no performance would be lost if perfect synchronization existed. For other values of $\tau$, D-OFDM1 DSTC is fairly robust. The partial performance loss is due to SNR reduction as previously explained. On the other hand, D-OFDM2 DSTC is completely robust against synchronization errors. For $\tau=0.5T_s$, D-OFDM2 DSTC performs as if it were perfectly synchronized. For $\tau=0.25 T_s$, D-OFDM2 DSTC performs slightly better than a conventional system with perfect synchronization. Similar results are obtained for the uncoded system over frequency-selective channels and hence they are not plotted for reasons of space.

Finally, to exploit the frequency diversity, we included channel coding and interleaving in a network with two relays. The $2\times 2$ orthogonal design and BPSK constellation are employed in the network. The $(2,1)$ and $(4,1)$ repetition codes and an interleaver of depth $D_{\intv}=10^4$ are utilized at the source to encode and interleave information bits. At the destination, soft outputs are computed from Eq.\eqref{eq:sym-decod1} and used for decoding. The simulation results are illustrated in Fig.~\ref{fig:M2_FS_R2_code}. The BER results of the uncoded system is also plotted for comparison. It can be seen that compared with an uncoded system, the slope of the curves and hence the diversity order is increased by the number of repetitions, i.e., 2 for a $(2,1)$ code and 4 for a $(4,1)$ code. The results show that the intrinsic diversity can be simply exploited using channel coding and interleaving.


\begin{figure}[h!]
\psfrag {double} [l] [] [0.8] {double sampling, $\tau=0.5T_s$}
\psfrag {tau05} [l] [] [0.8] {D-OFDM1, $\tau=0.5T_s$}
\psfrag {tau04} [l] [] [0.8] {D-OFDM1, $\tau=(0.4 \& 0.6)T_s$}
\psfrag {tau03} [l] [] [0.8] {D-OFDM1, $\tau=(0.3 \& 0.7)T_s$}
\psfrag {tau00} [l] [] [0.8] {D-OFDM1, $\tau=0 \& T_s$}
\psfrag {double05} [l] [] [.8] {D-OFDM2, $\tau=(0\&0.5\&1)T_s$}
\psfrag {double025} [l] [] [.8] {D-OFDM2, $\tau=0.25T_s$}
\psfrag {BER} [] [] [1] {BER}
\psfrag {Total Power} [t] [] [1] {$P/N_0$dB}
\centerline{\epsfig{figure={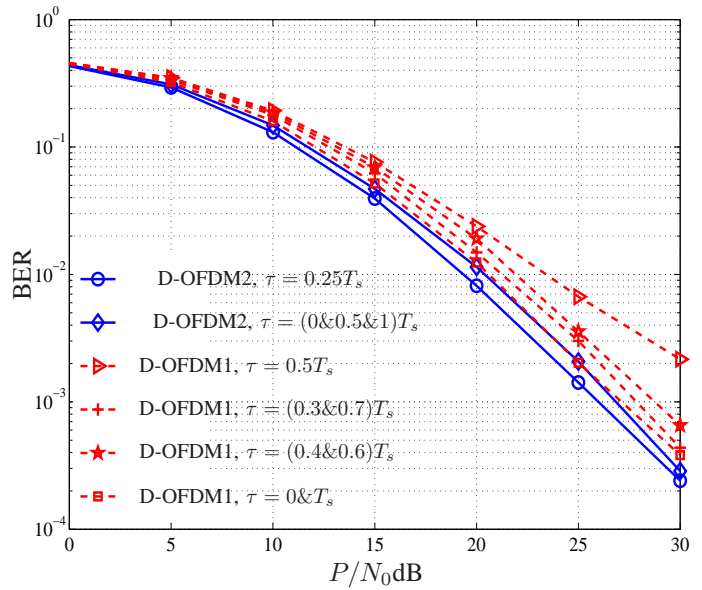},height=7.5cm,width=9cm}}
\caption{Simulation BER of a relay network with $R=2$ relays over frequency-selective channels, D-OFDM1 (symbol-rate sampling) and D-OFDM2 (double sampling), ($N=64, N_{\cp}=12$), using orthogonal designs and BPSK, $\tau_1=0, \tau_2=\tau$.}
\label{fig:M2_FS_R2}
\end{figure}

\begin{figure}[t]
\psfrag {double05} [l] [] [.8] {D-OFDM2, $\tau=(0\&0.5\&1)T_s$}
\psfrag {double025} [l] [] [.8] {D-OFDM2, $\tau=0.25T_s$}
\psfrag {tau05} [l] [] [0.8] {D-OFDM1, $\tau=0.5T_s$}
\psfrag {tau04} [l] [] [0.8] {D-OFDM1, $\tau=(0.4 \& 0.6)T_s$}
\psfrag {tau03} [l] [] [0.8] {D-OFDM1, $\tau=(0.3 \& 0.7)T_s$}
\psfrag {tau02} [l] [] [0.8] {D-OFDM1, $\tau=(0.2 \& 0.8)T_s$}
\psfrag {tau00} [l] [] [0.8] {D-OFDM1, $\tau=0 \& T_s$}
\psfrag {BER} [] [] [1] {BER}
\psfrag {Total Power} [t] [] [1] {$P/N_0$ (dB)}
\centerline{\epsfig{figure={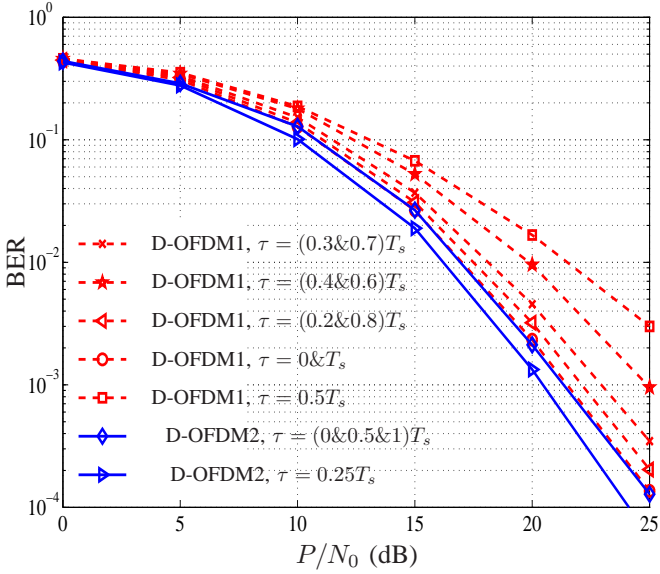},height=7.5cm,width=9cm}}
\caption{Simulation BER of a network with $R=4$ relays over frequency-selective channels, D-OFDM1 (symbol-rate sampling) and D-OFDM2 (double sampling), ($N=64, N_{\cp}=12$), using QOSTBC, $d_1=\tau_1=0$, $d_i$ is uniformly random, $\tau_i=\tau$ for $i=2,3,4$.}
\label{fig:M2_FS_R4}
\end{figure}

\begin{figure}[t]
\psfrag {fs-tau05-ds} [l] [] [.8] {D-OFDM2, $\tau=(0\&0.5\&1)T_s$}
\psfrag {fs-tau25-ds} [l] [] [.8] {D-OFDM2, $\tau=0.25T_s$}
\psfrag {fs-tau05} [l] [] [0.8] {D-OFDM1, $\tau=0.5T_s$}
\psfrag {fs-tau04} [l] [] [0.8] {D-OFDM1, $\tau=(0.4 \& 0.6)T_s$}
\psfrag {fs-tau03} [l] [] [0.8] {D-OFDM1, $\tau=(0.3 \& 0.7)T_s$}
\psfrag {fs-tau00} [l] [] [0.8] {D-OFDM1, $\tau=0 \& T_s$}
\psfrag {D-DSTC} [l] [] [0.8] {D-DSTC, $\tau=0$}
\psfrag {BLER} [b] [] [1] {BLER}
\psfrag {Total Power} [t] [] [1] {$P/N_0$ (dB)}
\centerline{\epsfig{figure={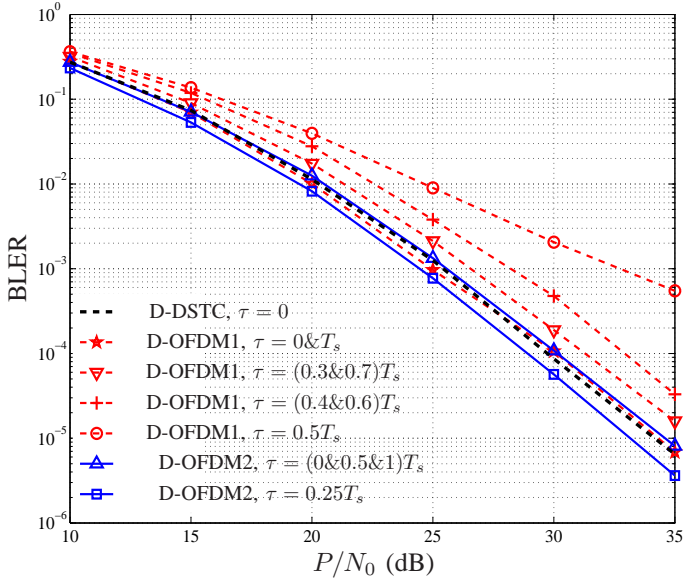},height=7.5cm,width=9cm}}
\caption{Simulation BLER of a network with $R=3$ relays, D-OFDM1 (symbol-rate sampling) and D-OFDM2 (double sampling), ($N=64, N_{\cp}=12$), D-DSTC \cite{D-DSTC-Y}, using circulant codes, $d_1=\tau_1=0$, $d_i$ is uniformly random, $\tau_i=\tau$ for $i=2,3$.}
\label{fig:M2_FS_R3}
\end{figure}

\begin{figure}[t]
\psfrag {CR1} [l] [] [1] {uncoded}
\psfrag {CR2} [l] [] [1] {coded (2,1)}
\psfrag {CR4} [l] [] [1] {coded (4,1)}
\psfrag {BER} [l] [] [1] {BER}
\psfrag {Total Power} [t] [] [1] {$P/N_0$ (dB)}
\centerline{\epsfig{figure={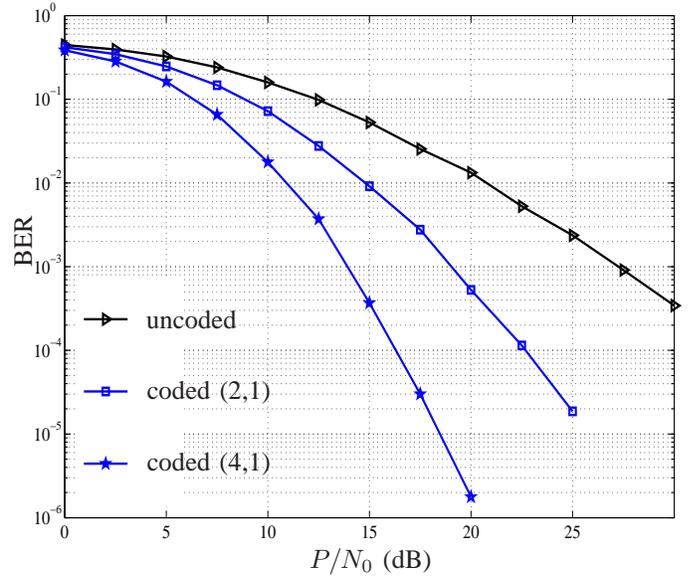},height=7.5cm,width=9cm}}
\caption{Simulation BER of a coded D-OFDM DSTC ($N=64, N_{\cp}=12$), $R=2$ relays over frequency-selective channels, using $2\times 2$ orthogonal designs and BPSK, $(2,1)$ and $(4,1)$ repetition codes and interleaver with $D_{\intv}=10^4$.}
\label{fig:M2_FS_R2_code}
\end{figure}

\section{Conclusion}
\label{sec:con}
While collecting channel information is challenging, especially in frequency-selective channels, synchronization errors are also inevitable in distributed space-time relay networks. Hence, in this paper a method was proposed that does not require any channel information and that nullifies synchronization errors. The method combines differential encoding and decoding with a frequency-domain approach to circumvent channel estimation and deal with synchronization errors. A new sampling scheme was developed to improve system performance against synchronization errors. It was shown through simulations that the method works well in both slow flat-fading and frequency-selective channels and for various synchronization error values.

\section*{Appendix: Derivation of \eqref{eq:y[n]-2}}

First, \eqref{eq:Xij} can be equivalently expressed as
\begin{equation}
\label{eq:Xij-2}
\begin{bmatrix}
X_{i,1}[m] \\
\vdots \\
X_{i,R}[m] 
\end{bmatrix}
= A \; \widehat{\B}_i 
\begin{bmatrix}
\widehat{Z}_{i,1}[m] \\
\vdots \\
\widehat{Z}_{i,R}[m]
\end{bmatrix}
\end{equation}
where
\begin{equation}
\left.
\begin{matrix}
\widehat{\B}_i=\B_i \\
\widehat{Z}_{i,r}[m]=Z_{i,r}[m]
\end{matrix}
\right\rbrace
\mbox{if} \; \C_i=\0,
\end{equation}
\begin{equation}
\left.
\begin{matrix}
\widehat{\B}_i=\C_i \\
\widehat{Z}_{i,r}[m]={Z}_{i,r}^*[\modN{-m}]
\end{matrix}
\right\rbrace
\mbox{if} \; \B_i=\0,
\end{equation}
for $i,r=1,\cdots,R$.

Thus, from \eqref{eq:yxG} and \eqref{eq:GnPn}
\begin{equation}
\label{eq:xd[n]}
\begin{split}
\x_i[n]=\begin{bmatrix}
x_{i,1}[n] \\
\vdots \\
x_{i,R}[n]
\end{bmatrix}=
\begin{bmatrix}
\DFT\{X_{i,1}[m] \} \\
\vdots \\
\DFT\{X_{i,R}[m] \} 
\end{bmatrix} \\
=A \widehat{\B}_i \begin{bmatrix}
\DFT \{ \widehat{Z}_{i,1}[m] \} \\
\vdots \\
\DFT \{ \widehat{Z}_{i,R}[m] \} 
\end{bmatrix}
\end{split}
\end{equation}
for $i=1,\cdots,R$.

Also, by taking the DFT of \eqref{eq:Zij} and \eqref{eq:CTR},
\begin{equation}
\label{eq:DFT_Rdi}
\DFT\{ Z_{i,r}[m]\}= \sqrt{P_0R} Q_i[n] s_r[n]+\psi_{i,r}[n],
\end{equation}

\begin{equation}
\label{eq:DFT_Rdi2}
\begin{split}
\DFT\{ {Z}_{i,r}^*[\modN{-m}]\}&=  \frac{1}{\sqrt{N}} \sum\limits_{m=0}^{N-1} Z_{i,r}^*[\modN{-m}] \et^{-j\frac{2\pi nm}{N}}\\&= \frac{1}{\sqrt{N}} \sum\limits_{k=0}^{N-1} Z_{i,r}^*[k] \et^{-j\frac{2\pi n \modN{-k}}{N}}\\&=
\frac{1}{\sqrt{N}} \sum\limits_{k=0}^{N-1} Z_{i,r}^*[k] \et^{j\frac{2\pi n k}{N}} \\&=
\IDFT\{ Z_{i,r}^*[k]\} =\left( \DFT\{Z_{i,r}[k]\} \right)^*\\&=\sqrt{P_0R} Q_i^*[n] s_r^*[n]+\psi_{i,r}^*[n]
\end{split}
\end{equation}
where $Q_i[n]=\sum \limits_{l=0}^{L-1} q_{i,l} \exp(-j2\pi nl/N)$ and $\psi_{i,r}[n]=\DFT\{\Psi_{i,r}[m]\}$. Note that in \eqref{eq:DFT_Rdi2} a variable change $k=\modN{-m}, m=\modN{-k}$ is performed. 

By substituting \eqref{eq:DFT_Rdi} into \eqref{eq:xd[n]} one has
\begin{equation}
\label{eq:xd[n]2}
\x_i[n]=A \sqrt{P_0R} \widehat{\B}_i \widehat{\s}_i[n] \widehat{Q}_i[n]+A \widehat{\B}_i \widehat{\psib}_i[n]
\end{equation}
where
\begin{equation}
\left.
\begin{matrix}
\widehat{\s}_i[n]=\s[n] \\
\widehat{Q}_{i}[n]=Q_{i}[n] \\
\widehat{\psib}_i[n]=\psib_i[n]
\end{matrix}
\right\rbrace
\mbox{if} \; \C_i=\0,
\end{equation}
\begin{equation}
\left.
\begin{matrix}
\widehat{\s}_i[n]=\s^*[n] \\
\widehat{Q}_{i}[n]=Q_{i}^*[n] \\
\widehat{\psib}_i[n]=\psib_i^*[n]
\end{matrix}
\right\rbrace
\mbox{if} \; \B_i=\0, 
\end{equation}
and 
\begin{equation*}
\begin{split}
\s[n]&=[\; s_1[n], \cdots, s_R[n] \;]^t,\\
\psib_i[n]&=[\; \psi_{i,1}[n], \cdots, \psi_{i,R}[n] \;]^t,
\end{split}
\end{equation*}
for $i=1,\cdots,R$.
Finally, we can derive \eqref{eq:y[n]-2} by substituting \eqref{eq:xd[n]2} into \eqref{eq:y[n]}.

\balance
\bibliographystyle{IEEEbib}
\bibliography{ref/references}

\end{document}